\documentclass[12pt]{iopart}

\usepackage{graphicx}
\usepackage{bm}
\usepackage{url}

\usepackage{iopams}

\usepackage{cite}

\begin{document}

\title{Generalizing the Sokolov-Ternov effect \\
       for radiative polarization in intense laser fields}

\author{X. S. Geng$^{1,2}$, 
        Z.G. Bu$^{1}$,
        Y.T. Wu$^{1,2}$, 
        Q.Q. Han$^{1,2}$, 
        C.Y. Qin$^{1,2}$， 
        W.Q. Wang$^{1,2}$, 
        X. Yan$^{1,2}$, 
        L.G. Zhang$^{1}$, 
        B. F. Shen$^{1,3,4}$, 
        L.L. Ji$^{1,3}$} 
\address{$^{1}$ State Key Laboratory of High Field Laser Physics, 
  Shanghai Institute of Optics and Fine Mechanics, 
  Chinese Academy of Sciences, Shanghai 201800, China}
\address{$^2$ University of Chinese Academy of Sciences, 
  Beijing 100049, China}
\address{$^3$ Center for Excellence in Ultra-intense Laser Science, 
  Chinese Academy of Sciences，Shanghai 201800, China}
\address{$^4$ Shanghai Normal University, Shanghai 200234, China}
\ead{
    \mailto{bfshen@mail.shcnc.ac.cn},
    \mailto{jill@siom.ac.cn}
    }

\begin{abstract}
    A consistent description of the radiative polarization for relativistic electrons in intense laser fields is derived by generalizing the Sokolov-Ternov effect in general field structure. 
    The new form together with the spin-dependent radiation-reaction force provides a complete set of dynamical equations for electron momentum and spin in strong fields. 
    When applied to varying intense fields, e.g. the laser fields, the generalized Sokolov-Ternov effect allows electrons to gain or lose polarization in any directions other than along the magnetic field in the rest frame of the electron. 
    The generalized theory is applied to the collision process between initially polarized/unpolarized high energy electrons with linearly polarized ultra-intense laser pulse, showing results that eliminate the dependence on specific choices of a quantization axis and spin initialization existing in spin-projection models. 
\end{abstract}

\maketitle 

\section{Introduction}
A charged particle with non-zero spin interacts with the external field via the Lorentz equation and the Stern-Gerlach force \cite{gerlach_experimentelle_1922}, 
where the particles can be deflected by the gradient of the magnetic field. 
For an ultra-relativistic electron in strong fields, when the field strength in its rest frame approaches the Schwinger limit \cite{heisenberg_folgerungen_1936}, 
$\gamma$-photon emission \cite{bamber_studies_1999} and its dependence on the electron’s spin state \cite{seipt_theory_2018} as well as the consequent spin-dependent radiation-reaction (RR) \cite{geng_spin-dependent_2020} come into play and alter the electron dynamics. 
On the other hand, the spin vector evolves according to the Thomas-Bargmann-Michel-Telegdi (T-BMT) equation \cite{thomas_motion_1926,thomas_i._1927,bargmann_precession_1959}. 
In the recent years, spin-relevant dynamics becomes a rising interest in the strong field regime \cite{seipt_theory_2018,geng_spin-dependent_2020,wen_spin-one-half_2017,del_sorbo_spin_2017,sorbo_electron_2018,li_ultrarelativistic_2019,seipt_ultrafast_2019,wu_polarized_2019} while the radiative polarization effect in varying fields, e.g. strong laser fields, is still not fully understood.

The spontaneous polarization of relativistic electrons in the static magnetic field due to the asymmetric spin-flip rates during photon emission is well known as the Sokolov-Ternov (S-T) effect \cite{sokolov_synchrotron_1968}. 
By modelling the process with spin-flips of an ensemble of electrons in the spin-up/-down states in the static magnetic field, the S-T effect successfully interprets the polarization evolution in storage rings.
This effect has been well confirmed and utilized to generate polarized leptons in conventional accelerator facilities \cite{baier_radiative_1972,camerini_measurement_1975,learned_polarization_1975,schwitters_azimuthal_1975,belomesthnykh_observation_1984}. 
The S-T effect was transplanted to a rotating standing wave recently \cite{del_sorbo_spin_2017}, where the magnetic field axis is constant and polarization can build up co-axial to the magnetic field within a few laser periods.
In more general cases, the field intensity and direction of, e.g. femtosecond laser pulse, vary within femtoseconds level and micron-meter scale. 
The situation becomes even more complicated in laser-driven plasmas such that spin evolution via radiative process can no longer be described by the S-T effect in long-term static or quasi-static fields. 

For more general consideration, the spin-flip probability rate of electrons in strong laser field, i.e. the Volkov state \cite{wolkow_uber_1935}, was calculated under the locally constant field approximation \cite{seipt_theory_2018}.
For numerical simulations in complex laser fields or laser-plasma interactions, a vector shall be derived from the polarization density matrix of the electron, so that sequential spin precession in external fields becomes tractable. 
The spin-flip probability given in \cite{seipt_theory_2018} allows the spin vector to be projected onto itself, i.e. it either flips or stays unchanged during the photon emission, namely the s-projection. 
Alternatively,  the spin vector of the test particle can be projected onto the magnetic axis in the rest frame of the electron after a $\gamma$-photon emission \cite{li_ultrarelativistic_2019,seipt_ultrafast_2019} according to the proposed, namely, the B-projection. 
Both approaches converge when the initial spin orientations are parallel to the magnetic field in the rest frame of the electrons $\mathbf{B}_{\rm{rest}}$. 
Projecting the spin state onto a chosen base after each photon emission event suffers from information loss along other possible basis that might be of measuring interest. 
It can also lead to substantial variation of the spin evolution in complex fields. 
For example, in case of initial spins perpendicular to $\mathbf{B}_{\rm{rest}}$, spins remain perpendicular during photon emission in the s-projection scenario and no polarization is built up along the $\mathbf{B}_{\rm{rest}}$. 
On the other hand, in the B-projection scenario the initial polarization vanishes quickly after one single emission. 
The issue caused by B-projection has been mitigated recently by adjusting the chosen axis on a measuring-of-interest basis \cite{li_polarized_2020}.
It should be noted that if the time scale of the change of the direction of $\mathbf{B}_{\rm{rest}}$ is larger than the time scale between photon emissions, the B-projection could be a good approximation of the radiative polarization effect.

In this article, we generalize the S-T effect in varying strong fields such that electron beams can gain polarization in a certain direction but do not lose polarization in other directions. 
We construct a self-consistent description of the spin dynamics in the strong field, based on the spin-dependent probability rates given by the quantum-electrodynamics (QED) calculations. 
The new approach preserves the full information of spin states after photon emission, thus capable of handling general field structures where no preferred projection basis is present. 
The generalized S-T model can be implemented into the particle-in-cell simulation to account for the polarization effect in laser-plasma interactions.

This paper is organized as follows: 
in section \ref{sec1}, we generalize the S-T effect to an arbitrary axis and relevant spin dynamics and numerical methods are presented;
in section \ref{sec2}, we compare the generalized S-T approach to the ones with a chosen set of projection axis.

\section{Theoretical models \label{sec1}}
\subsection{The generalized Sokolov-Ternov effect}
In the Sokolov-Ternov effect, electrons get polarized along the magnetic field through asymmetric spin-flip process where the electrons tend to be anti-parallel to the magnetic field. The polarization process is dominated by the equation

\begin{equation}\label{eq:equilibrium}
\frac{\rmd}{\rmd t}N^{\uparrow}=\frac{\rmd P^{\downarrow\uparrow}}{\rmd t}N^{\downarrow}-\frac{\rmd P^{\uparrow\downarrow}}{\rmd t}N^{\uparrow}
\end{equation}

where $N^{\uparrow},N^{\downarrow}$ denote the number of electrons in the spin-up/-down state along the chosen axis, e.g. the magnetic field direction in the S-T effect, and $P$ denotes the probability rate of the transition between the spin-up/-down states in a photon emission. 
The polarization along the chosen axis is defined by $\langle\mathbf{s}\rangle=Pol=\frac{N^{\uparrow}-N^{\downarrow}}{N^{\uparrow}+N^{\downarrow}}$ and we have the equation of polarization

\begin{equation}\label{eq:polAB}
    \frac{\rmd}{\rmd t}Pol+(A+B)Pol=A-B
\end{equation}

where $A=\frac{\rmd P^{\downarrow\uparrow}}{\rmd t},B=\frac{\rmd P^{\uparrow\downarrow}}{\rmd t}$ are defined for convenience. The solution for the equation is
\begin{equation}\label{eq:Polt}
    Pol\left(t\right)=\frac{A-B}{A+B}\left[1-\exp{\left(-\frac{t}{\tau}\right)}\right]+{\rm{Pol}}_0\exp{\left(-\frac{t}{\tau}\right)}
\end{equation}
where $\tau=\frac{1}{A+B}$ is the polarization time and ${\rm{Pol}}_0$ is the initial polarization along the chosen axis.

In general, the above solution applies to any polarization axis other than the axis of magnetic field as long as the spin flip rates $A$ and $B$ are known from a given theory. 
As an example, we show in the following how a full solution is derived based on the radiative spin-flip probability rate in general field. 
We employ the calculation via the spin density matrix method  \cite{seipt_theory_2018}
\begin{equation}\label{eq:fliprate}
    \frac{\rmd P_{\rm{flip}}^s}{\rmd\psi \rmd\delta}=
    -\frac{\alpha}{2b}\frac{\delta^2}{1-\delta}
    \left\{\frac{{\rm{Ai}}^\prime\left(z\right)}{z}
    -s_\zeta\frac{\rm{Ai}\left(z\right)}{\sqrt{z}}
    -s_\kappa^2\left[\rm{Ai}_1\left(z\right)+\frac{{\rm{Ai}}^\prime\left(z\right)}{z}\right]\right\}
\end{equation}
where $\psi$ is the phase, 
$\delta$ the energy fraction of the emitted photon to the electron energy, 
$\alpha$ is the fine structure constant, 
$b=\left(\hbar k^\mu\cdot p_\mu\right)/m^2c^2$ is the energy parameter with 
$\hbar$ being the reduced Planck constant, 
$m$ the electron mass, 
$c$ the speed of light in vacuum, 
$k^\mu$ and $p^\mu$ the four vectors of the wave-vector and the electron momentum, 
$z=\left[\frac{\delta}{\left(1-\delta\right)\chi_e}\right]^{2/3}$, 
$\chi_e=e\hbar/m^3c^4\left|F^{\mu\nu}\cdot p_{\mu\nu}\right|$ is the nonlinear quantum parameter with 
$e$ being the fundamental charge and 
$F^{\mu\nu}$ the electro-magnetic tensor
and $g=1+\frac{\delta^2}{2\left(1-\delta\right)}$. 
Here $\rm{Ai}$ \& ${\rm{Ai}}^\prime$ are the Airy function and its first order derivative and $\rm{Ai}_1\left(y\right)=\int_{y}\rm{Ai}\left(x\right)dx$. 
$s_\zeta$ \& $s_\kappa$ are the components of spin vector on $\mathbf{B}_{\rm\rm{{res}t}}$ and $\mathbf{k}_{\rm{\rm{rest}}}$ (the wave vector of the field in the rest frame). 
The spin-flip probability was derived based on the locally constant-crossed field approximation (LCFA) that treats the external field as constant and crossed during, i.e. $\mathbf{E}\cdot\mathbf{B}=0$, the QED-photon emission process. 
It is applicable when $a_0^3/\chi_e\gg1$ \cite{baier_quantum_1989,dinu_quantum_2016,di_piazza_implementing_2018} where $a_0=\frac{eE_0}{\omega mc}$ is the invariant field strength of the laser field 
with $E_0$ being the electric field strength and $\omega$ the angular frequency of the laser field. 

Following Eq. (\ref{eq:fliprate}), the spin-flip rates along the magnetic field $\zeta$, electric field $\eta$ and the wave vector $\kappa$ in the rest frame are defined as a mutually orthogonal and complete set of bases, 
which is for analytical convenience, valid under $a_0\ll\left({mc^2}/{\hbar\omega}\right)$($\approx4700$ for typical Ti:Sa lasers).
The flip rates in their pure states ($\left|\mathbf{s}\right|=1$) along those directions are
\numparts
\begin{eqnarray}\label{eqs:AB}
    A^\zeta=\int{-\frac{\alpha}{2b}\frac{\delta^2}{1-\delta}\left\{\frac{{\rm Ai}^\prime\left(z\right)}{z}+\frac{\rm Ai\left(z\right)}{\sqrt z}\right\}d\delta\cdot\frac{d\psi}{dt}}\label{eq:Azeta}\\
    B^\zeta=\int{-\frac{\alpha}{2b}\frac{\delta^2}{1-\delta}\left\{\frac{{\rm Ai}^\prime\left(z\right)}{z}-\frac{\rm Ai\left(z\right)}{\sqrt z}\right\}d\delta\cdot\frac{d\psi}{dt}}\label{eq:Bzeta}\\
    A^\eta=B^\eta=\int{-\frac{\alpha}{2b}\frac{\delta^2}{1-\delta}\left\{\frac{{\rm Ai}^\prime\left(z\right)}{z}\right\}d\delta\cdot\frac{d\psi}{dt}}\label{eq:ABeta}\\
    A^\kappa=B^\kappa=\int{-\frac{\alpha}{2b}\frac{\delta^2}{1-\delta}\left\{\frac{{\rm Ai}^\prime\left(z\right)}{z}-\left[\rm Ai_1\left(z\right)+\frac{{\rm Ai}^\prime\left(z\right)}{z}\right]\right\}d\delta\cdot\frac{d\psi}{dt}}\label{eq:ABkappa}
\end{eqnarray}
\endnumparts

\begin{figure}
    \centering
    \includegraphics[width=1\textwidth]{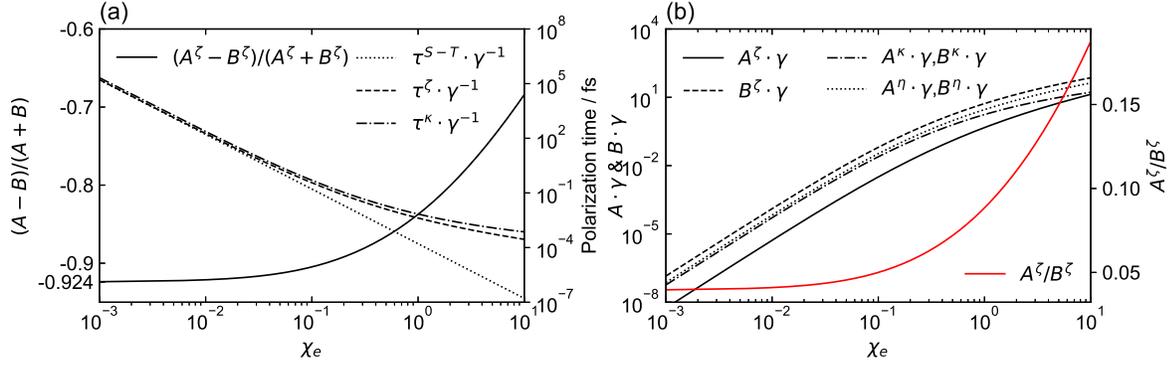}
    \caption{
        (a) (left axis) The polarization limit ${\left(A^\zeta-B^\zeta\right)}/{\left(A^\zeta+B^\zeta\right)}$ (solid). 
        (right axis) The polarization time $\tau={1}/{\left(A+B\right)}$ divided by $\gamma$ along the $\zeta$ direction (dashed), the classical approximation of the S-T effect (dotted) and polarization time in the $\kappa$ direction (dotted-dashed). 
        The polarization time in the $\eta$ direction equals that in the $\zeta$ direction.
        (b) (left axis) The $A$s and $B$s in Eqs. (\ref{eq:Azeta}-\ref{eq:ABkappa}) scaled by $\gamma$ and (right axis) $A^\zeta/B^\zeta$ (red solid).
    }
    \label{fig1}
\end{figure}

The above spin-flip rates indicate that the electron can only get polarized along the direction $\zeta$ (the magnetic field). 
Combining it with Eq. (\ref{eq:Polt}) one can find that the electrons can only get depolarized along other directions where we have $A=B$. 
Eqs. (\ref{eq:Polt}) and (\ref{eq:Azeta}-\ref{eq:ABkappa}) together describe the evolution of polarization due to photon emission. 
In the classical limit that $\chi_e\ll1$, our theory reproduces the well-known polarization limit in the S-T effect, i.e. ${\left(A-B\right)}/{\left(A+B\right)}$, of -0.924 and the polarization time scale of S-T effect $\tau^{\rm{S-T}}=\frac{8\sqrt{3}}{15}\frac{\hbar^2}{mce^2}\gamma\chi_e^{-3}$ \cite{ternov_synchrotron_1995} as shown in Figure \ref{fig1}(a).
The polarization time in the $\kappa$ direction $\tau^\kappa$ is also shown and one can find that $\tau^\eta=\tau^\zeta$ as $A^\eta+B^\eta=A^\zeta+B^\zeta$. 
Disparity emerges when $\chi_e$ is approaching and going beyond unity, where the polarization time is larger than that of the S-T effect and the polarization limit declines ($\left|\frac{A/B\ -1}{A/B\ +1}\right|$ declines when $A/B$ increases), as suggested by the red line in Figure \ref{fig1}(b).

In Eq. (\ref{eq:fliprate}) we consider the flip rates of a unit spin vector $\mathbf{s}$ which is the quantum average of the spin state \cite{mane_spin-polarized_2005}. 
Following that, the evolution of polarization along each axis can be derived according to Eq. (\ref{eq:Polt}) using the flip rates of $\mathbf{s}$ along those axes in Eq. (\ref{eq:fliprate}), 
which is equivalent to the evolution of the mixed state of spin. 
By dealing with the polarization in a complete set of basis, the full spin information of an electron after radiation is correctly preserved therefore the results are insensitive to the choice of projection axis or initial spin orientations as they are for the s-projection and B-projection. 

Apart from the radiative polarization effect, the spin vector, as well as the polarization, undergoes precession around an axis in the rest frame of the electron according to the T-BMT equation (see \cite{jackson_classical_1998})
\begin{equation}\label{eq:TBMT}
    \frac{\rmd}{\rmd t}\mathbf{s}=\frac{e}{mc}\mathbf{s}\times
    \left[
        \left(a_e+\frac{1}{\gamma}\right)\mathbf{B}
        -a_e\frac{\gamma}{\gamma+1}\left(\mathbf{\beta}\cdot\mathbf{B}\right)\mathbf{\beta}
        -\left(a_e+\frac{1}{\gamma+1}\right)\mathbf{\beta}\times\mathbf{E}
    \right]
\end{equation}
where $a_e\approx1.16\times{10}^{-3}$ \cite{hanneke_cavity_2011} is the anomalous magnetic moment of electron \cite{schwinger_quantum-electrodynamics_1948}, $\mathbf{B}$ is the magnetic field, $\mathbf{E}$ is the electric field, $\mathbf{\beta}=\mathbf{v}/c$ is the normalized velocity. 
Eq. (\ref{eq:Polt}) and (\ref{eq:TBMT}) offer a complete description of the spin dynamics. In our case, we consider precession of the polarization vector by replacing the spin vector s with $\mathbf{Pol}\equiv\langle\mathbf{s}\rangle$. 

\subsection{Spin-dependent radiation-reaction}
The dynamics becomes self-consistent when one includes the spin-dependent radiation-reaction. 
For the polarization $\langle\mathbf{s}\rangle$, the radiation probability rate is \cite{seipt_theory_2018}
\begin{equation}\label{eq:radiation_rate}
    \frac{\rmd P^s}{\rmd\psi\rmd\delta}=-\frac{\alpha}{b}
    \left\{
        \rm{Ai}_1(z)+g\frac{2\rm Ai'(z)}{z} + \langle\mathbf{s}\rangle_\zeta\delta\frac{\rm{Ai}(z)}{\sqrt{z}}
    \right\}
\end{equation}
By averaging the radiated photon energy using Eq. (\ref{eq:radiation_rate}),
\begin{equation}\label{eq:average_energy}
    \frac{\rmd{\bar{\delta}}^s}{\rmd\psi}=\int{\delta\frac{\rmd P^s}{\rmd \psi \rmd\delta}\rmd\delta}
\end{equation}
RR force is thus the recoil of the averaged energy loss from photon emission by momentum conservation. As shown by Eq. (\ref{eq:radiation_rate}), radiation energy has a negative dependence on the magnetic component and thus the electrons with negative spin component along $\mathbf{B}_{\rm{rest}}$, i.e. in the anti-parallel state, tends to radiate more energy and experience strong RR force.

We would also like to mention that another spin-dependent force, the Stern-Gerlach force induced by the gradient of the magnetic field $\sim\nabla\left(\mathbf{s}\cdot\mathbf{B}\right)$, is also present in strong laser fields.
However, this effect is weak in the considered regime \cite{wen_spin-one-half_2017} when compared to the spin-dependent RR effect \cite{geng_spin-dependent_2020}. The latter is $10^4$ times stronger than the former in terms of the deflection angle between the oppositely polarized test electrons in collision with ultra-intense laser pulses. 
Therefore, the Stern-Gerlach effect is not included in the following calculations.

\subsection{Numerical methods}
The theoretical model based on Eq. (\ref{eq:Polt}), (\ref{eq:TBMT}) and (\ref{eq:radiation_rate}) is a complete set of description for spin-dependent electron dynamics.
Following the standard semi-classical semi-QED calculation \cite{gonoskov_extended_2015}, the particle motion between each photon emission event is treated classically through the Lorentz equation ${\rmd\mathbf{p}}/{\rmd t}=-e(\mathbf{E}+\mathbf{v}\times\mathbf{B})$. 
The RR effect is considered by adding a RR force to the electron, which is achieved by losing momentum of  ${\bar{\delta}}^s\cdot\sqrt{\gamma^2-1}mc^2\cdot\frac{\rmd \psi}{\rmd t}\Delta t$ at each time step where $\Delta t$ is the time step. 
Since we consider polarization, each test electron is assigned with a polarization property to represent the average of the spin vectors for an ensemble of electron spin states. 
The evolution of the polarization is calculated by Eq. (\ref{eq:Polt}), (\ref{eq:Azeta}-\ref{eq:ABkappa}) and (\ref{eq:TBMT}).

For comparison considerations we also perform QED Monte-Carlo (MC) simulations of RR \cite{gonoskov_extended_2015} and radiative spin-dynamics \cite{geng_spin-dependent_2020} for the s-projection and B-projection scenarios. 
Here each test electron is assigned with a spin vector that either flips along itself (s-projection) according to the flip rate Eq. (\ref{eq:fliprate}) or gets projected onto the $\mathbf{B}_{\rm{rest}}$(B-projection) according to the probability rate of projection $\mathbf{s}$ to $\pm{\hat{\mathbf{B}}}_{\rm{rest}}$ \cite{seipt_ultrafast_2019}. 
In the meantime, the spin vector undergoes precession according to the T-BMT equation. 
The original S-T formula is also included in the MC calculation for comparison. 

While our model takes QED average for the photon emission, for an electron beam with many particles or laser-driven plasmas, the average in each phase space cell can be a good representation of those treated stochastically \cite{geng_quantum_2019,harvey_quantum_2017}

\section{Results \label{sec2}}
\begin{figure}
    \centering
    \includegraphics[width=1\textwidth]{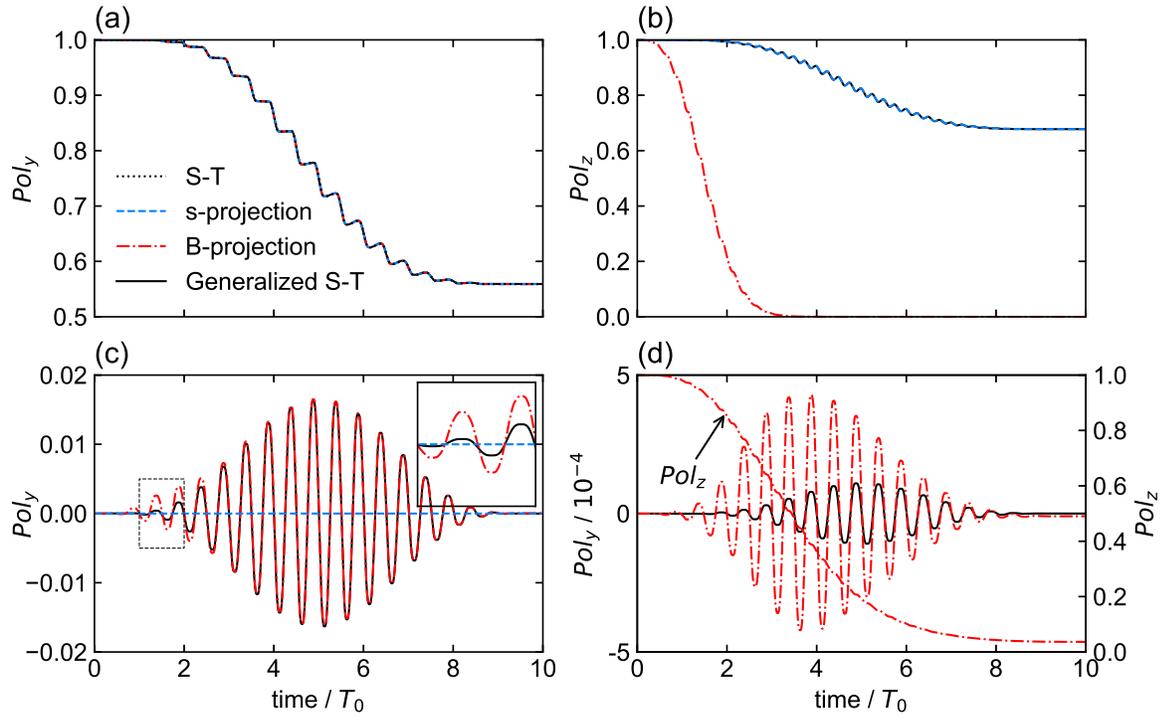}
    \caption{
        The polarization during the collision with a LP laser pulse. 
        (a) $Pol_y$ of the electrons initially polarized along the magnetic field of the LP laser ($y$-axis). 
        The polarization $Pol_y$ of the generalized S-T model (black solid), the MC results of the S-T formula (black dashed), s-projection model (blue dotted) and B-projection model (red dotted-dashed) coincide with each other. 
        (b) $Pol_z$ and 
        (c) $Pol_y$ of the electrons initially polarized along the wave vector of the laser ($z$-axis). 
        The solid-lined box in (c) is the area in the zoomed dashed-lined box. 
        (d) Results of (b) and (c) for $a_0=10$.
    }
    \label{fig2}
\end{figure}

We then apply our approach to the collision between relativistic electrons and intense laser pulses to evaluate the polarization effect. 
First, we benchmark our results with the QED-MC simulation by colliding a linearly polarized laser with electrons of $\gamma_0=1000$ polarized along the magnetic field direction of the laser, e.g. y-axis, as shown in Figure \ref{fig2}(a). The LP laser is approximated by 
\begin{equation}
    \mathbf{E}=\hat{\mathbf{x}}E_x=\hat{\mathbf{x}}E_0\exp{\left(-\frac{x^2+y^2}{w_0^2}\right)}\cos{\left(\psi\right)}\cos^2{\left(\frac{\psi}{2N}\right)}
\end{equation}
and $\mathbf{B}=\hat{\mathbf{y}}E_x/c$, where $E_0$ is the electric amplitude corresponding to $a_0=100$, $w_0=5\lambda$ is the beam waist, 
$\psi$ is the laser phase and 
$N=20$ is the full length of the pulse in the unit of wavelength (800nm). 
To explicitly compare the polarization evolution without being disturbed by other effects, we first switch off RR and spin precession in the comparison to avoid stochastic effects induced by MC, w
hich is important when $\chi_e>0.1$ \cite{vranic_quantum_2016,harvey_quantum_2017,geng_quantum_2019}.
One can clearly see that the polarization evolution of our energy averaged modelling agrees with those of the MC results. 
This is because the initial polarization orientation converges with the magnetic field $\mathbf{B}_{\rm{rest}}$ in the rest frame of the electron. 
The polarization decreases every half period and slightly increases every other half period because the magnetic field of the laser is oscillating during the collision. 
Eventually the polarization declines to about 60\%, which is consistent with the known S-T effect that electron spins tend to be aligned towards the negative direction of the magnetic field as suggested by $A^\zeta<B^\zeta$ in Figure \ref{fig1}(b).

For further comparison, evolution for the longitudinal polarized electrons are shown in Figure \ref{fig2}(b-d). 
One can see that while the projection models are inconsistent with each other, our theory agrees with the s-projection in terms of $Pol_z$ and the B-projection for $Pol_y$, 
i.e., it preserves the information of longitudinal polarization $Pol_z$ and in the meantime predicts polarization build-up $Pol_y$ along the B-field. 
The s-projection model is unable to gain any polarization in the y-direction (Figure \ref{fig2}(c)) and the B-projection model predicts a fast depolarization of $Pol_z$ (Figure \ref{fig2}(b)). 
The transverse polarization is consistent for the B-projection and the generalized S-T modelling in general. However, substantial mismatch at the beginning of the interaction emerges (see the solid-lined box in Figure \ref{fig2}(c)). 
It originates from the projection process where spin vectors rapidly fall onto the y-axis. 
After all spin vectors become coaxial to the y-axis in a few laser periods consistency is recovered. 
Such mismatch can be more distinct when spin vectors cannot be fully turned into parallel to the y-axis, e.g., when the field strength is weaker such that not all electrons radiate. 
As shown in Figure \ref{fig2}(d) for $a_0=10$, where the LCFA still holds as $a_0^3/\chi_e\gg1$, the non-zero $Pol_z$ after collision indicates that during the process there are non-radiating electrons and their spin vectors remains undisturbed (stay longitudinal). 
Significant discrepancy is seen for the $Pol_y$, where the B-projection overestimates the polarization build-up as compared to the generalized S-T model. 
The above comparisons illustrate that the projection modelling of individual electron spins could lead to loss of information and/or substantial error for electrons in intense electromagnetic fields.

\begin{figure}
    \centering
    \includegraphics[width=1\textwidth]{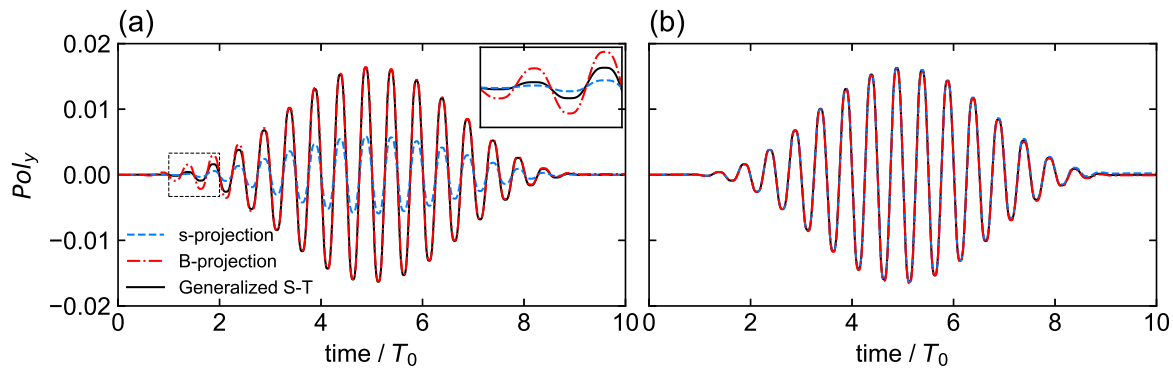}
    \caption{
        The transverse polarization $Pol_y$ of unpolarized electrons during the collision with a LP laser pulse for different type of initialization of the unpolarized electrons. The polarization of the generalized S-T models is set to 0. 
        (a) Spin vectors of the test electrons in the MC models are uniformly scattered to the $4\pi$ space or 
        (b) equally distributed along the y-axis. The solid-lined box in (a) is the zoomed area in the dashed-lined box.
    }
    \label{fig3}
\end{figure}

We now apply our theory to electrons with zero initial polarization. 
In the s- or B-projection modelling, since each electron carries a unit spin vector, initialization of unpolarized electrons becomes subtle. 
One may scatter the spin vectors of the test electrons uniformly into $4\pi$ solid angles to generate overall zero polarization as suggested in \cite{li_ultrarelativistic_2019}. 
Alternatively, electrons can be separated equally into two groups of opposite direction along a chosen axis as suggested by the S-T effect. 
Both initializations are capable of modelling the unpolarized state in the MC algorithm, but no discussion has been seen on the consequences. 
In the modelling of the generalized S-T effect, we simply set $Pol=0$ for the electron, which is equivalent to scattering the $Pol$s into $4\pi$ solid space or grouping into opposite sets because the dependence on initial polarization is canceled for opposite $Pol$s (see Eq. (\ref{eq:Polt})) in terms of the radiative polarization effect. 
The results for the $4\pi$ initialization are shown in Figure \ref{fig3}(a) where the B-projection model agrees well with the generalized S-T theory in terms of $Pol_y$ (along the B-field). 
When spin vectors are equally distributed along the y-axis, all approaches converge, as shown in Figure \ref{fig3}(b). 
In the unpolarized cases, all the modellings gain no polarization along $Pol_x$ and $Pol_z$ as electrons can only get polarized along the $\mathbf{B}_{\rm{rest}}$ direction and the zero results are not displayed here. 
For initialization along the $z$-axis, the result can be generalized from the $Pol_y$ in Figure \ref{fig2}(b), which is the summation of oppositely polarized cases where the B-projection predicts consistent result with the generalized S-T model and s-projection gains no $Pol_y$. 
One notices the diversion in Figure \ref{fig3}(a) between the B-projection modelling and the generalized S-T. 
Again, this is due to the projection process where electron spin vectors rapidly fall onto y-axis, as demonstrated in Figure \ref{fig2}(c) and \ref{fig2}(d). 
Furthermore, the $4\pi$ initialization forbids the s-projection model to gain more degree of polarization than the cases where the spin vectors are paraxial to the $\mathbf{B}_{\rm{rest}}$ axis as they can only flip or stays unchanged along itself.

The results indicate that while previous models depend on specifically chosen initial conditions, e.g. spin orientations with respect to the chosen projection axis, the generalized S-T model considering polarization is consistent in handling any initial conditions.

\begin{figure}
    \centering
    \includegraphics[width=1\textwidth]{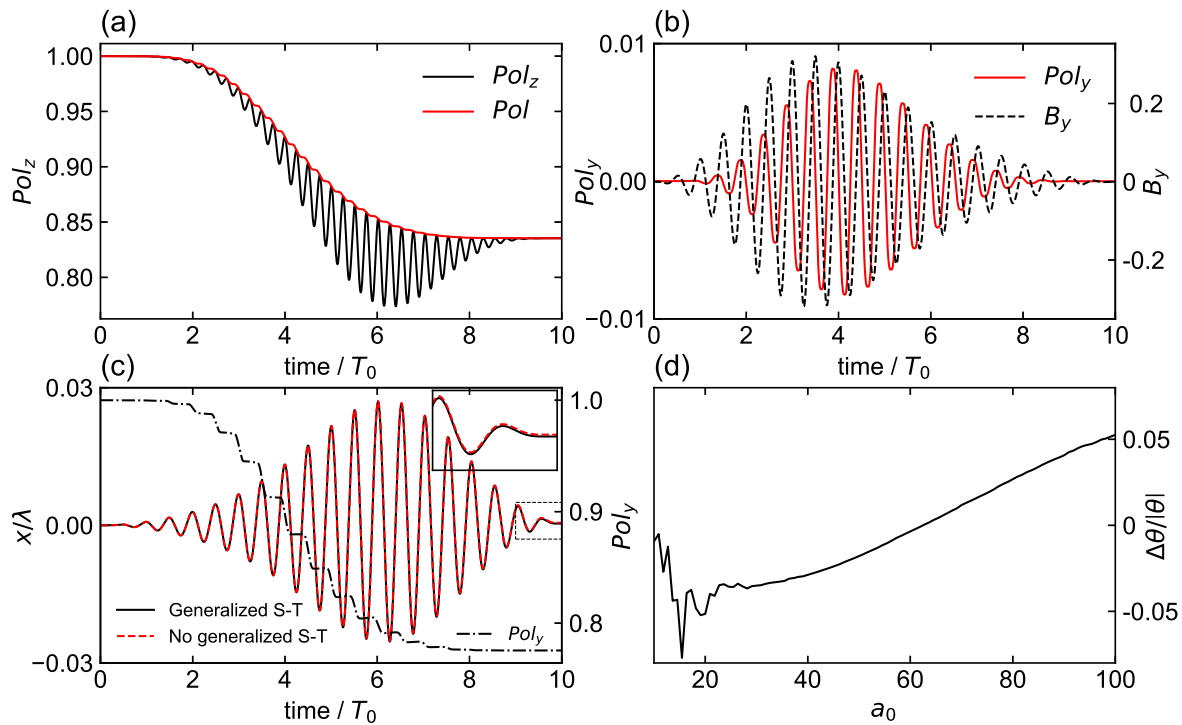}
    \caption{
        Collision between LP laser of $a_0=100$ and an electron of $\gamma_0=1000$. 
        (a) The longitudinal polarization $Pol_z$ and the total polarization $Pol$ of an electron initially longitudinally polarized along $z$-axis. 
        (b) (left axis) The evolution of $Pol_y$ and (right axis) $B_y$ in the rest frame. 
        (c) (left axis) The oscillation amplitude of electrons of $Pol_y=1$ (black solid). The dashed red line corresponds to cases without radiative depolarization effect. The lines in the solid box are the zoomed details in the dashed box. (right axis) The evolution of $Pol_y$ (dashed-dotted). 
        (d) The ratio between the deflection angle difference and the deflection angle in (c) for different field strength.
    }
    \label{fig4}
\end{figure}

We now include spin precession, spin-dependent RR and the generalized S-T effect to investigate the collision between electrons of $\gamma_0=1000$ and LP laser of $a_0=100$ in Figure \ref{fig4}. 
In Figure \ref{fig4}(a) and \ref{fig4}(b) the electron is initially longitudinally polarized along $z$-axis where the $Pol_z$ component drastically oscillates due to spin precession and the total polarization gradually decreases due to radiative depolarization effect similar to that in Figure \ref{fig2}(b). 
The depolarization effect is weaker here due to the energy loss and consequent smaller $\chi_e$ during the collision. 
On the other hand, the transverse polarization $Pol_y$ oscillates with a smaller amplitude as compared to that in Figure \ref{fig2}(c) due to the energy loss from radiation. 
It should be noted that the evolution of ${Pol}_y$ in this case is identical to that of an unpolarized electron of $Pol=0$. 
Considering the spin-dependent RR effect, we evaluate the spin-dependent deflection effect including the radiative depolarization, as shown in Figure \ref{fig4}(c), 
where the black/red line corresponds to the transverse location $x$ for electron of $Pol_y=1$ with/without radiative depolarization. 
One can see that the radiative depolarization effect slightly depresses the deflection effect because $Pol_y$ gradually decreases as shown by the dotted-dashed line. 
The effect of radiative depolarization on the deflection angle is evaluated in a larger range for $a_0=10\sim100$ in Figure \ref{fig4}(d). 
The disparity is subtle in this case but can be magnified in, for example, the flat-top pulse that is discussed in \cite{geng_spin-dependent_2020} and possibly the laser-plasma interactions.

\section{Conclusions}
In conclusion, we derived the Sokolov-Ternov-like polarization effect and generalized it to arbitrary directions other than along the magnetic field. We compare the generalized Sokolov-Ternov model to the previous models that require specific choices of the projection of spin vectors and find that our model is capable of handling the radiative polarization effect in any direction without the need for a specific projection axis. By considering the spin-dependent radiation-reaction, the spin precession and the spin polarization effect, we present a consistent description of the spin dynamics in the intense laser field.

\ack
The authors would like to thank Dr. Sateesh Mane for very insightful discussion. This work is supported by the National Science Foundation of China (Nos. 11875307 and 11935008) and the Strategic Priority Research Program of Chinese Academy of Sciences (Grant No. XDB16010000).


\section*{References}
\begin{thebibliography}{10}

    \bibitem{gerlach_experimentelle_1922}
    Walther Gerlach and Otto Stern.
    \newblock Der experimentelle {Nachweis} der {Richtungsquantelung} im
      {Magnetfeld}.
    \newblock {\em Z. Physik}, 9(1):349--352, December 1922.
    
    \bibitem{heisenberg_folgerungen_1936}
    W.~Heisenberg and H.~Euler.
    \newblock Folgerungen aus der {Diracschen} {Theorie} des {Positrons}.
    \newblock {\em Z. Physik}, 98(11):714--732, November 1936.
    
    \bibitem{bamber_studies_1999}
    C.~Bamber, S.~J. Boege, T.~Koffas, T.~Kotseroglou, A.~C. Melissinos, D.~D.
      Meyerhofer, D.~A. Reis, W.~Ragg, C.~Bula, K.~T. McDonald, E.~J. Prebys, D.~L.
      Burke, R.~C. Field, G.~Horton-Smith, J.~E. Spencer, D.~Walz, S.~C. Berridge,
      W.~M. Bugg, K.~Shmakov, and A.~W. Weidemann.
    \newblock Studies of nonlinear {QED} in collisions of 46.6 {GeV} electrons with
      intense laser pulses.
    \newblock {\em Phys. Rev. D}, 60(9):092004, October 1999.
    
    \bibitem{seipt_theory_2018}
    D.~Seipt, D.~Del~Sorbo, C.~P. Ridgers, and A.~G.~R. Thomas.
    \newblock Theory of radiative electron polarization in strong laser fields.
    \newblock {\em Phys. Rev. A}, 98(2):023417, August 2018.
    
    \bibitem{geng_spin-dependent_2020}
    X.~S. Geng, L.~L. Ji, B.~F. Shen, B.~Feng, Z.~Guo, Q.~Q. Han, C.~Y. Qin, N.~W.
      Wang, W.~Q. Wang, Y.~T. Wu, X.~Yan, Q.~Yu, L.~G. Zhang, and Z.~Z. Xu.
    \newblock Spin-dependent radiative deflection in the quantum radiation-reaction
      regime.
    \newblock {\em New J. Phys.}, 22(1):013007, January 2020.
    
    \bibitem{thomas_motion_1926}
    L.~H. Thomas.
    \newblock The {Motion} of the {Spinning} {Electron}.
    \newblock {\em Nature}, 117(2945):514--514, April 1926.
    
    \bibitem{thomas_i._1927}
    L.~H. Thomas.
    \newblock I. {The} kinematics of an electron with an axis.
    \newblock {\em The London, Edinburgh, and Dublin Philosophical Magazine and
      Journal of Science}, 3(13):1--22, January 1927.
    
    \bibitem{bargmann_precession_1959}
    V.~Bargmann, Louis Michel, and V.~L. Telegdi.
    \newblock Precession of the {Polarization} of {Particles} {Moving} in a
      {Homogeneous} {Electromagnetic} {Field}.
    \newblock {\em Phys. Rev. Lett.}, 2(10):435--436, May 1959.
    
    \bibitem{wen_spin-one-half_2017}
    Meng Wen, Christoph~H. Keitel, and Heiko Bauke.
    \newblock Spin-one-half particles in strong electromagnetic fields: {Spin}
      effects and radiation reaction.
    \newblock {\em Phys. Rev. A}, 95(4):042102, April 2017.
    
    \bibitem{del_sorbo_spin_2017}
    D.~Del~Sorbo, D.~Seipt, T.~G. Blackburn, A.~G.~R. Thomas, C.~D. Murphy, J.~G.
      Kirk, and C.~P. Ridgers.
    \newblock Spin polarization of electrons by ultraintense lasers.
    \newblock {\em Phys. Rev. A}, 96(4):043407, October 2017.
    
    \bibitem{sorbo_electron_2018}
    D.~Del Sorbo, D.~Seipt, A.~G.~R. Thomas, and C.~P. Ridgers.
    \newblock Electron spin polarization in realistic trajectories around the
      magnetic node of two counter-propagating, circularly polarized, ultra-intense
      lasers.
    \newblock {\em Plasma Phys. Control. Fusion}, 60(6):064003, April 2018.
    
    \bibitem{li_ultrarelativistic_2019}
    Yan-Fei Li, Rashid Shaisultanov, Karen~Z. Hatsagortsyan, Feng Wan, Christoph~H.
      Keitel, and Jian-Xing Li.
    \newblock Ultrarelativistic {Electron}-{Beam} {Polarization} in {Single}-{Shot}
      {Interaction} with an {Ultraintense} {Laser} {Pulse}.
    \newblock {\em Phys. Rev. Lett.}, 122(15):154801, April 2019.
    
    \bibitem{seipt_ultrafast_2019}
    Daniel Seipt, Dario Del~Sorbo, Christopher~P. Ridgers, and Alec G.~R. Thomas.
    \newblock Ultrafast {Polarization} of an {Electron} {Beam} in an {Intense}
      {Bi}-chromatic {Laser} {Field}.
    \newblock {\em arXiv:1904.12037 [hep-ph, physics:physics]}, April 2019.
    \newblock arXiv: 1904.12037.
    
    \bibitem{wu_polarized_2019}
    Yitong Wu, Liangliang Ji, Xuesong Geng, Qin Yu, Nengwen Wang, Bo~Feng, Zhao
      Guo, Weiqing Wang, Chengyu Qin, Xue Yan, Lingang Zhang, Johannes Thomas, Anna
      H\"utzen, Markus B\"uscher, T.~Peter Rakitzis, Alexander Pukhov, Baifei Shen,
      and Ruxin Li.
    \newblock Polarized electron-beam acceleration driven by vortex laser pulses.
    \newblock {\em New J. Phys.}, 21(7):073052, July 2019.
    
    \bibitem{sokolov_synchrotron_1968}
    A.~A. Sokolov and I.M. Ternov.
    \newblock {\em Synchrotron {Radiation}}.
    \newblock Akademie, Berlin, 1968.
    
    \bibitem{baier_radiative_1972}
    V.~N. Baier.
    \newblock {RADIATIVE} {POLARIZATION} {OF} {ELECTRONS} {IN} {STORAGE} {RINGS}.
    \newblock {\em Sov. Phys. Usp.}, 14(6):695, 1972.
    
    \bibitem{camerini_measurement_1975}
    U.~Camerini, D.~Cline, J.~Learned, A.~K. Mann, and L.~K. Resvanis.
    \newblock Measurement of the radiative electron polarization in a 2.4-{GeV}
      storage ring.
    \newblock {\em Phys. Rev. D}, 12(7):1855--1858, October 1975.
    
    \bibitem{learned_polarization_1975}
    J.~G. Learned, L.~K. Resvanis, and C.~M. Spencer.
    \newblock Polarization of colliding ${e}^{+}{e}^{\ensuremath{-}}$ beams at
      spear ii.
    \newblock {\em Phys. Rev. Lett.}, 35(25):1688--1690, December 1975.
    
    \bibitem{schwitters_azimuthal_1975}
    R.~F. Schwitters, A.~M. Boyarski, M.~Breidenbach, F.~Bulos, G.~J. Feldman,
      G.~Hanson, D.~L. Hartill, B.~Jean-Marie, R.~R. Larsen, D.~L\"uke, V.~L\"uth,
      H.~L. Lynch, C.~C. Morehouse, J.~M. Paterson, M.~L. Perl, T.~P. Pun,
      P.~Rapidis, B.~Richter, W.~Tanenbaum, F.~Vannucci, F.~M. Pierre, G.~S.
      Abrams, W.~Chinowsky, C.~E. Friedberg, G.~Goldhaber, J.~A. Kadyk, A.~M.
      Litke, B.~A. Lulu, B.~Sadoulet, G.~H. Trilling, J.~S. Whitaker, F.~C.
      Winkelmann, and J.~E. Wiss.
    \newblock Azimuthal asymmetry in inclusive hadron production by
      ${e}^{+}{e}^{\ensuremath{-}}$ annihilation.
    \newblock {\em Phys. Rev. Lett.}, 35(20):1320--1322, November 1975.
    
    \bibitem{belomesthnykh_observation_1984}
    S.~A. Belomesthnykh, A.~E. Bondar, M.~N. Yegorychev, V.~N. Zhilitch, G.~A.
      Kornyukhin, S.~A. Nikitin, E.~L. Saldin, A.~N. Skrinsky, and G.~M. Tumaikin.
    \newblock An observation of the spin dependence of synchrotron radiation
      intensity.
    \newblock {\em Nuclear Instruments and Methods in Physics Research Section A:
      Accelerators, Spectrometers, Detectors and Associated Equipment},
      227(1):173--181, November 1984.
    
    \bibitem{wolkow_uber_1935}
    D.~M. Wolkow.
    \newblock {\"Uber} eine {Klasse} von {L\"osungen} der {Diracschen} {Gleichung}.
    \newblock {\em Z. Physik}, 94(3):250--260, March 1935.
    
    \bibitem{li_polarized_2020}
    Yan-Fei Li, Rashid Shaisultanov, Yue-Yue Chen, Feng Wan, Karen~Z.
      Hatsagortsyan, Christoph~H. Keitel, and Jian-Xing Li.
    \newblock Polarized ultrashort brilliant multi-gev $\ensuremath{\gamma}$ rays
      via single-shot laser-electron interaction.
    \newblock {\em Phys. Rev. Lett.}, 124(1):014801, January 2020.
    
    \bibitem{baier_quantum_1989}
    V.~N. Baier, V.~M. Katkov, and V.~M. Strakhovenko.
    \newblock Quantum radiation theory in inhomogeneous external fields.
    \newblock {\em Nucl. Phys. B}, 328(2):387--405, December 1989.
    
    \bibitem{dinu_quantum_2016}
    Victor Dinu, Chris Harvey, Anton Ilderton, Mattias Marklund, and Greger
      Torgrimsson.
    \newblock Quantum {Radiation} {Reaction}: {From} {Interference} to
      {Incoherence}.
    \newblock {\em Phys. Rev. Lett.}, 116(4):044801, January 2016.
    
    \bibitem{di_piazza_implementing_2018}
    A.~Di~Piazza, M.~Tamburini, S.~Meuren, and C.~H. Keitel.
    \newblock Implementing nonlinear {Compton} scattering beyond the
      local-constant-field approximation.
    \newblock {\em Phys. Rev. A}, 98(1):012134, July 2018.
    
    \bibitem{ternov_synchrotron_1995}
    I.~M. Ternov.
    \newblock Synchrotron radiation.
    \newblock {\em Phys.-Usp.}, 38(4):409, 1995.
    
    \bibitem{mane_spin-polarized_2005}
    S.~R. Mane, Yu~M. Shatunov, and K.~Yokoya.
    \newblock Spin-polarized charged particle beams in high-energy accelerators.
    \newblock {\em Rep. Prog. Phys.}, 68(9):1997--2265, August 2005.
    
    \bibitem{jackson_classical_1998}
    J.~D. Jackson.
    \newblock {\em Classical {Electrodynamics} 3rd edition}.
    \newblock Wiley, New York, 1998.
    
    \bibitem{hanneke_cavity_2011}
    D.~Hanneke, S.~Fogwell~Hoogerheide, and G.~Gabrielse.
    \newblock Cavity control of a single-electron quantum cyclotron: {Measuring}
      the electron magnetic moment.
    \newblock {\em Phys. Rev. A}, 83(5):052122, May 2011.
    
    \bibitem{schwinger_quantum-electrodynamics_1948}
    Julian Schwinger.
    \newblock On {Quantum}-{Electrodynamics} and the {Magnetic} {Moment} of the
      {Electron}.
    \newblock {\em Phys. Rev.}, 73(4):416--417, 1948.
    
    \bibitem{gonoskov_extended_2015}
    A.~Gonoskov, S.~Bastrakov, E.~Efimenko, A.~Ilderton, M.~Marklund, I.~Meyerov,
      A.~Muraviev, A.~Sergeev, I.~Surmin, and E.~Wallin.
    \newblock Extended particle-in-cell schemes for physics in ultrastrong laser
      fields: {Review} and developments.
    \newblock {\em Phys. Rev. E}, 92(2):023305, August 2015.
    
    \bibitem{geng_quantum_2019}
    X.~S. Geng, L.~L. Ji, B.~F. Shen, B.~Feng, Z.~Guo, Q.~Yu, L.~G. Zhang, and
      Z.~Z. Xu.
    \newblock Quantum reflection above the classical radiation-reaction barrier in
      the quantum electro-dynamics regime.
    \newblock {\em Commun Phys}, 2(1):66, June 2019.
    
    \bibitem{harvey_quantum_2017}
    C. N. Harvey, A.~Gonoskov, A.~Ilderton, and M.~Marklund.
    \newblock Quantum {Quenching} of {Radiation} {Losses} in {Short} {Laser}
      {Pulses}.
    \newblock {\em Phys. Rev. Lett.}, 118(10):105004, March 2017.
    
    \bibitem{vranic_quantum_2016}
    Marija Vranic, Thomas Grismayer, Ricardo~A. Fonseca, and Luis~O. Silva.
    \newblock Quantum radiation reaction in head-on laser-electron beam
      interaction.
    \newblock {\em New J. Phys.}, 18(7):073035, July 2016.
    
    \end{thebibliography}
\end{document}